# Hand Posture's Effect on Touch Screen Text Input Behaviors: A Touch Area Based Study

Christopher Thomas
Department of Computer Science
University of Pittsburgh
5428 Sennott Square
210 South Bouquet Street
Pittsburgh, PA 15260
chris@cs.pitt.edu

Brandon Jennings
Department of Computer Engineering
University of Pittsburgh
1140 Benedum Hall
3700 O'Hara St
Pittsburgh, PA 15213
bbj5@pitt.edu

**ABSTRACT**
Mobile devices with touch keyboards have become ubiquitous, but text entry on these devices remains slow and errorprone. Understanding touch patterns during text entry could be useful in designing robust error-correction algorithms for soft keyboards. In this paper, we present an analysis of text input behaviors on a soft QWERTY keyboard in three different text entry postures: index finger only, one thumb, and two thumb. Our work expands on the work of [1] by considering the entire surface area of digit contact with the smartphone keyboard, rather than interpreting each touch as a single point. To do this, we captured touch areas for every key in a lab study with 8 participants and calculated offsets, error rates, and size measurements. We then repeated the original experiment described in [1] and showed that significant differences exist when basing offset calculations on touch area compared to touch points for two postures.

**INTRODUCTION**
Despite advances in error correction algorithms, text entry on small touch screen devices is slow and error prone [8]. The all too well-known problems of occlusion, fat finger, and the lack of haptic feedback contribute to the poor accuracy on such devices. Even so, the number of touch screen devices considers to grow rapidly, with a projected 3 billion touch screen devices to in 2016 [16]. Furthermore, these keyboards are increasingly being designed for direct finger input, rather than requiring users to use a stylus [1].

To address the high-error rates encountered on these devices, sophisticated error correction algorithms have been designed which take into account learned models of typical text entry behaviors. For instance, the model may take into account the fact that for certain keys, users' fingers are likely to land slightly offset from the center of the key. This information can be useful when translating the user's key press location to the intended key on the keyboard. Unfortunately, many factors impact keyboarding behaviors, such as the hand posture of the user holding the device and how they are entering text on it. For instance, text entry behaviors are different when holding the device with one hand and using only one thumb when compared to simply typing with the index finger one character at a time. In this work, we consider three different postures and develop a model of text entry behaviors for each.

To develop our posture-based models, we ran a pilot study with 8 participants who were instructed to enter a variety of sentences using our custom data collection application. We log the touched area of the screen for each key press. Using this data, we determine error rates, sizes, and offsets for typical key presses for each key. Here, offsets refer to the distance from the center of the key the user was intending to press to each pixel value covered by the user's touch area. We consider both vertical and horizontal offsets in this study. Though Azenkot et al. have already performed a similar study, their models are based only on touch points: the center of the area touched by the user. In reality, however, the user's finger does not hit the screen at exactly one point. Rather, the finger contact actually forms a touch area, the 2-d region consisting of all locations in which the user's finger contacts the screen. Our hypothesis in this paper is that touch point analysis may be an oversimplification, which discards information which may impact the overall model and its conclusions. Thus, rather than relying on a simplification of the user's input, we use all available information when constructing our models.

The remainder of this paper is structured as follows. We first provide an overview of the relevant related work to this study. We then provide a description of our data collection program and our user study. Afterwords, we present data visualizations constructed from the user study data and analyze and compare our model to the one constructed using the method of Azenkot et al. Finally, we present some overall conclusions of our study and some ideas for future work.

**RELATED WORK**
This work is is an extension of [1] and builds upon the growing body of literature of touch screen interfaces and text input. There are three basic areas strongly related to our current study: exploration of general touch performance on capacitive touch screens, touch performance during text entry, and improving techniques for key detection and auto-correction algorithms, to which our work provides a foundation for.

**Touch Performance**

There have been several papers investigating general touch performance, though not in the context of text entry as it involves higher cognitive and motor skills than more simple pointing tasks.

Lee Zhai [12] compared soft and hard buttons with varying parameters such as finger vs. stylus, auditory and tactilevibrato feedback and button size, and presented a vast array of findings.

Henze et al. [7] designed a game in which users were to hit targets across a screen, and analyzed their touch offsets. What was observed was that user touch events were systematically skewed to just left and just above the bottom-right corner of the screen, suggesting that touch events are shifted towards the a position where the user's thumb would naturally touch the screen if the phone is held in the right hand.

In addition to finding touch events consistently offset from their intended targets, Holz and Baudisch [9] also showed that users perceived contact points to be about the center of the fingernail, which is above the actual contact point along the finger's axis.

Wang and Ren [17] conducted a study to explore different human finger input properties such as contact area, contact shape, and contact orientation of all five fingers. Their results showed that the five fingers of a single hand exhibit different abilities and potentials for target selection. Notably, the index, middle, and ring finger are more precise than the thumb and pinky fingers.

**Text Entry Performance**

Various papers have aimed to better understand and improve the performance of text entry on touch screen devices.

Findlater et al. [4] used large touch surfaces to show that personalized input models for ten-finger typing greatly improves key-pressed classification accuracy over a generic model designed for any user. Findlater and Wobbrock [3] go on to provide empirical evidence for the benefit of such personalization, both in terms of typing performance and user experience via keyboard adaptation.

A model was presented by MacKenzie and Zhang [14] based on Fitt's Law for predicting expert speed on pen-based soft keyboards. MacKenzie and Soukoreff [13] later proposed a two-thumb entry model that is empirically validated and adjusted by Clarkson et al. [2]. Both models are capable of predicting the speed of expert typists while neglecting error patterns.

**Key Detetction and Auto-Correct**

One method of key detection and auto-correction is dynamically changing the underlying key size of soft keyboards depending on context.

Goodman et al. [5] uses a combination of touch point distribution and character probabilities given by natural language models to correct user error. For example, if a user hits 'q' and then hits 'i', it is most probable that the user intended to hit 'u' because 'u' is next to 'i' and it almost always couples with 'q'. They also showed that language models can significantly reduce error rates by a factor of 1.67 and 1.87.

Gunawardana et al. [6] proposed a less aggressive key-target method which provided a robust input method that does not prevent users from typing their desired text. They found through empirical evaluation that using anchored dynamic key-targets significantly reduced key-stroke error compared to the state-of-the-art. In future work, Rudchenko et al. [15] introduced a texting game that generates ideal training data for key-target resizing, in addition to improving user experience by providing target practice and improvement highlights.

Kristensson and Zhai [11] proposed a geometric pattern matching method to word level error correction. In their method, the hit points on a stylus keyboard can be matched against patterns formed by letter key center positions of legitimate words in a lexicon.

Although there is a relatively new error tolerant method of text entry, the gesture keyboard [10][18][19], that can be linked with touch keyboards [11], we are interested in touch area only for the purposes of this study.

**EXPERIMENTAL DESIGN**

Our experiment followed a within-subjects design, with posture being the within-subjects factor. Subjects were men and women between the ages of 25-65, all of whom were righthanded. Following the method used by [1], subjects were instructed to type "as accurately and as naturally as possible." All subjects completed the experiment while seated using the same Samsung Galaxy Note II. The study consisted of typing sentences on a custom keyboard application. The application displayed randomly generated sentence phrases for the user to re-enter and a QWERTY keyboard with minimal functionality. The application was designed to capture the fundamental behavior of users on a soft keyboard without potentially distracting design features (such as a backspace key).

The keyboard dimensions on the data collection application were marginally smaller than those of the default Android keyboard. There is no backspace to remove mistyped characters as it would influence the natural error rate a user might exhibit. Furthermore, there are no numbers, symbols, or punctuation.

**DATA ANALYSIS**

Our data analysis follows in the spirit of [1]. We calculate three metrics (error rate, offset, and size of touch area) from the user study for each of the postures. As with Azenkot, we first clean the data by removing outliers more than 1.5 times the key height away from their target key center. For instance, if the user was to enter a "t" and his touch was on the letter "a" the touch is removed from the training set. However, if the user had hit a "y" instead, the touch would have been preserved. The error rate is defined as the number of incorrect

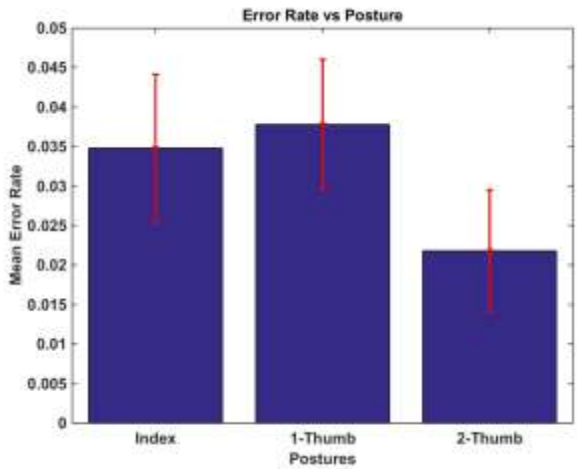

Figure 1. Error Rates for Each Posture Type

Figure 1 shows our results for error rates. The whiskers on the figure mark the 95% confidence intervals for the mean error rates. Post-hoc analysis revealed that the following postures were significantly different with $p < 0.05$: index finger vs key presses divided by the total number of key presses. However, our error rate calculation is presumably identical to that of Azenkot et al. as the underlying API translates the touch to the key press. We do not try to change the key that the API reports was pressed (by considering our touch area model), though we believe that doing so could in principle yield a lower error rate. While taking our learned model into account when translating touches to key presses in comparison with the default Android behavior would be interesting but is beyond the scope of this study.

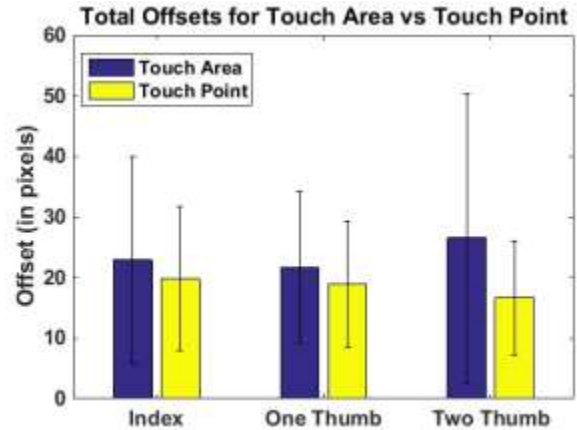

Figure 2. Overall Offsets for Each Posture Type

We define offset as the distance for each pixel in the touch area to the center of the button the user was attempting to press. First, we determine the touch area of the user's touch and all X-Y pixel locations within the touch area. Next, we calculate the horizontal and vertical distances for each pixel separately (to separate horizontal offsets from vertical offsets). We compute these offsets for each key on the keyboard and visualize both the horizontal and vertical offsets. We also compute the euclidean distance from each X-Y location to the center of the key and compute an average total offset for each posture (which is the average of offsets from all the keys on the keyboard). We use this aggregate offset calculation to determine whether posture has a significant impact on offset.

To determine whether or not posture had a significant impact on each of the metrics, we computed each subject's scores for each of the three postures. We then ran a repeated-measures ANOVA with posture as the within-subject factor. All analysis was performed with a 95% confidence threshold. We confirmed that for all three metrics (error rate, offset, and size) that posture had a significant impact. For each metric, we then combined all subjects into one overarching posture category and performed post-hoc analysis using a 2-sample t-test between each posture to determine which postures were significantly different from each other.

two thumbs and one thumb vs two thumbs. Interestingly, our error-rate differences mirror those of Azenkot et al., who also found significant differences between error rates between the same postures. Our tests revealed somewhat lower error rates than Azenkot et al., but the relative error rates are similar; both studies place 1 thumb input at the highest error rate, followed by index finger, followed by 2 thumb input with the lowest error rate. This was a surprising result for me (as well as for Azenkot) because text entry with users only using their index finger intuitively seems more accurate. Similarly, input with two thumbs seems like it may be the most error-prone, but in fact it is the least error-prone posture. We posit that one-thumb input is likely error-prone because of the reaching that must occur across the screen (to the left side). This notion could be tested by determining the per-key error rate and verifying if the errors are higher on the left side. Similarly, we hypothesize that index text entry may be so error-prone because moving from key to key involves the subject moving his or her entire hand up then over then back down, whereas one thumb and two thumb modalities fix the hand in position and only allow the digit to move (i.e. there is an additional degree of freedom with the index text input).

Next, we compare total euclidean offsets computed using the method of [1] compared to using the entire touch area for each of the postures. We show the results of these calculations in Figure 2. Touch area based offsets showed significant differences $p < 0.05$ for the index finger vs two thumb and one thumb vs two thumb comparisons. We then performed twosample t-test comparisons between the touch area and touch point conditions, comparing overall offsets within each posture. We observed that one thumb and two thumb overall offsets were significantly different when touch area was used as a basis for computation when compared to touch point. We further decomposed the problem into horizontal and vertical offsets and computed both for touch point vs. touch area offsets to determine where the differences between the two offset calculations lied. We determined that no significant horizontal offset differences existed. However, in both the one thumb and two thumb postures, the vertical offsets calculated from the touch area were significantly different than those calculated using touch points (with the vertical offsets almost always larger when touch area was used).

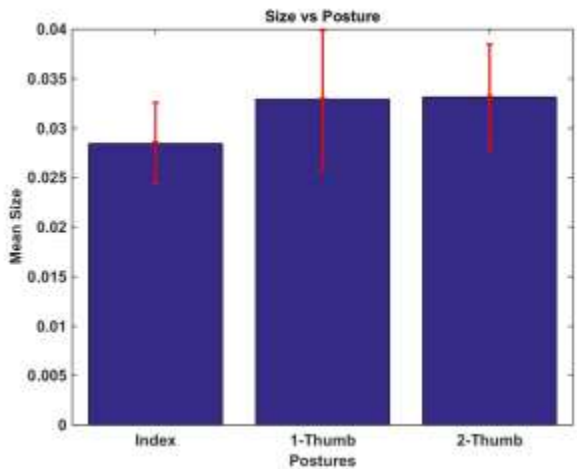

Figure 3. Overall Size for Each Posture Type

Figure 3 shows a comparison of the average size of each touch area for the three postures. Post-hoc analysis revealed significant differences with $p < 0.01$ between index finger vs one thumb and index finger vs two thumbs. No significant difference in touch area size between the one thumb and two thumb postures was observed. Intuitively this result makes sense because the index finger is narrower than the thumbs, which resulted in it having smaller touch areas. The fact that the sizes of one thumb input compared to two thumb input were not significantly different suggests that the overall touch area size between these two postures was not significantly different. One may have originally attempted to explain the lower error rate of the two thumb posture by suggesting that the one thumb posture resulted in a larger area of the thumb making contact with the screen (when stretching for instance). However, this analysis reveals that that this not the case. This shows that the difference in error rates between the postures is not a question of *how* the thumbs contact the screen differently in the two postures but rather a question of *where* the thumbs land. This suggests that there are fundamental differences in offsets between the two postures which cannot be explained away by claiming it is a difference of touch area size.

**Touch Visualizations**

To illustrate the differences between the touch area and touch point methods, we constructed data visualizations in the same manner as Azenkot et al. To do this, for each key we generate a list of X-Y locations touched for that key. As with Azenkot, we consider the *goal* key rather than the key actually pressed (excepting far outliers who are not removed from the dataset). Thus, if a user is to enter a "t" and his key press registers a "y" we still consider these touch points as belonging to "t" rather than "y." For touch area, each touch results in only one touch point being added to the key's list. For touch area, all points within the touch area are added to the list. We then compute the 95% confidence ellipses around the data points and visualize our results. Figures 4 and 5 show the confidence ellipses constructed using the touch point and touch area methods, respectively. A small glyph appears next to each visualization illustrating the posture.

The confidence ellipse visualizations reveal several interesting differences between the touch point and touch area methods. However, for the most part, the two visualizations are very similar, with the touch area ellipses simply appearing wider. One of the observations of Azenkot et al. was that keys in the middle tended to have the least overlapping ellipses compared to those on the far right or left. The touch point study tends to reiterate this finding. However, the touch area ellipses reveal that the situation is more complex and significantly more overlap between neighboring ellipses is occurring. Thus, the touch point method tends to *underestimate* the amount of overlap actually occurring. For all three postures, touch area ellipses revealed much more overlap than those computed using the touch point method. For the most part however, the angle of the ellipses remained roughly the same. Many of the findings of Azenkot et al. were visible in both visualizations, such as the right offset of the key press on the space key and the tendency of the space key to not be entirely utilized. These findings suggest that while many of the conclusions drawn from Azenkot et al.'s work are valid, error-correction model attempting to take overlapping touch regions into account should consider basing their models on touch area.

**OFFSET VISUALIZATIONS**

Continuing in the style of Azenkot et al., we compute visualizations of the vertical and horizontal offsets (computed as described above) for each letter on the keyboard. Figures 6 and 7 illustrate horizontal offsets computed using touch points and touch areas respectively. Figures 8 and 9 illustrate vertical offsets for touch points and touch areas respectively. Recall from our previous discussion that no horizontal offsets were shown to be significantly different when comparing touch point vs. touch area offsets per letter. However, vertical offsets for both the one thumb and two thumb postures computed using touch areas demonstrated statistically significant differences $p < 0.05$ when compared to their touch point offset counterparts. In most cases, the vertical offsets were observed to be greater in the touch area offset bank. These differences suggest error-correction models taking into account vertical offsets should consider testing models relying on touch area as well as those involving touch points. However, we do note that for the most part the overall direction of the offsets computed using the touch area method correspond exactly to those of the touch point method (there are only a few minor differences in shading). In fact, a side by side comparison of the touch point vs. touch offset models reveals only minor differences in shading for the majority of the letters. We believe this is mostly because the elliptical touch area's sides cancel each other out (that is, the right side of the ellipse cancels out relative offsets on the left side, etc.), thus causing the central touch point to be a relatively good approximation of the overall offset of the touch.

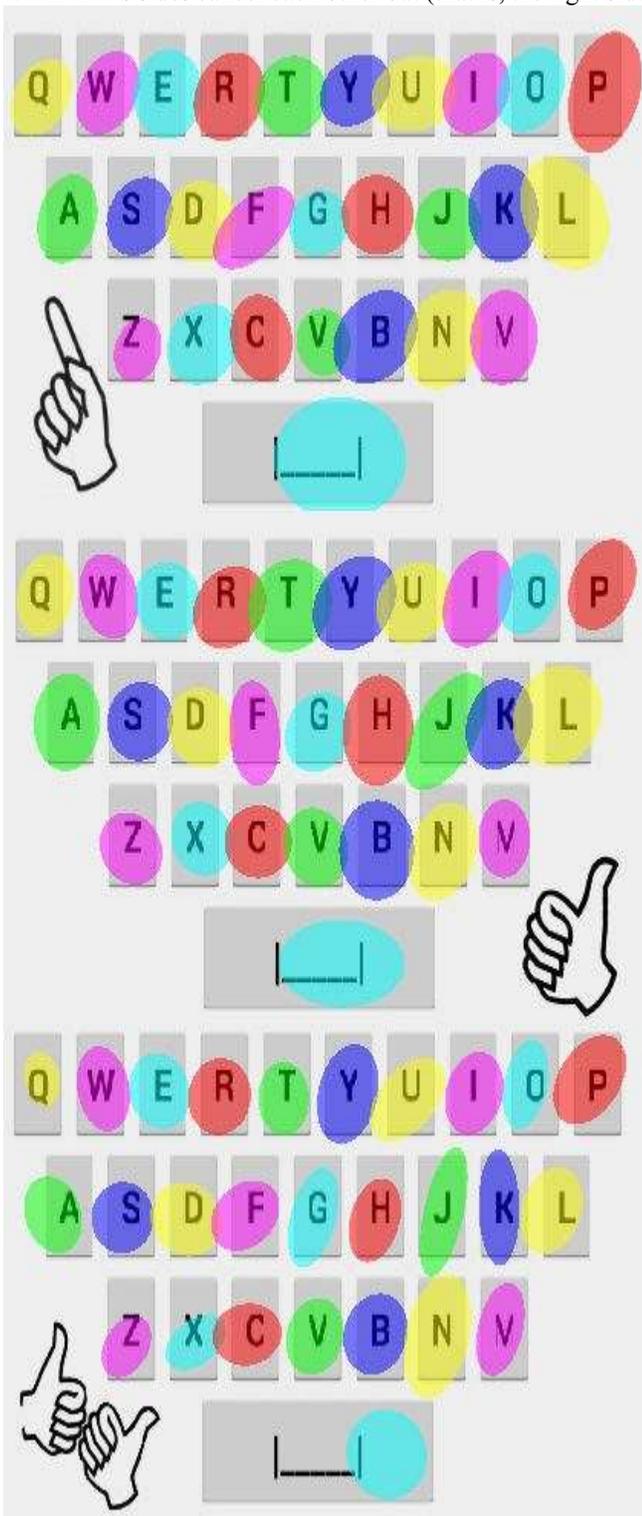

Figure 4. Confidence Ellipses for Touch Point Method

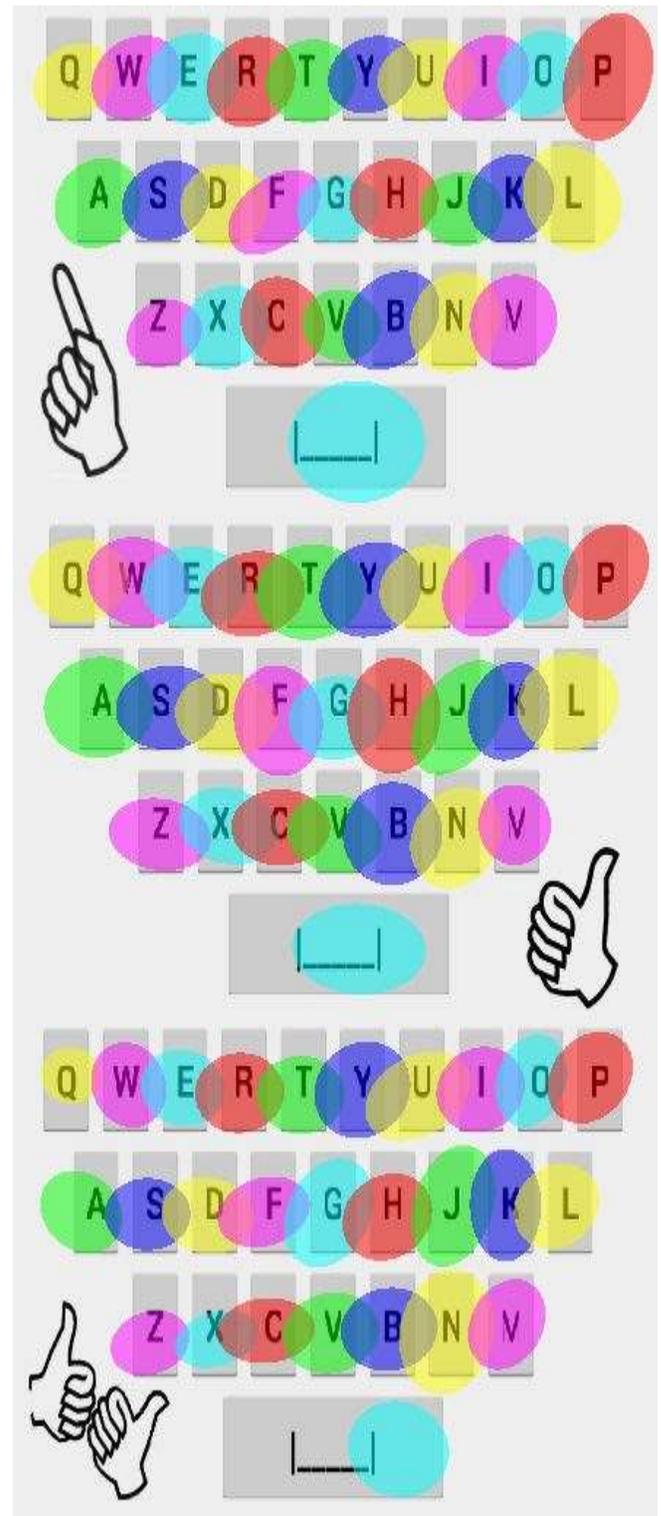

Figure 5. Confidence Ellipses for Touch Area Method

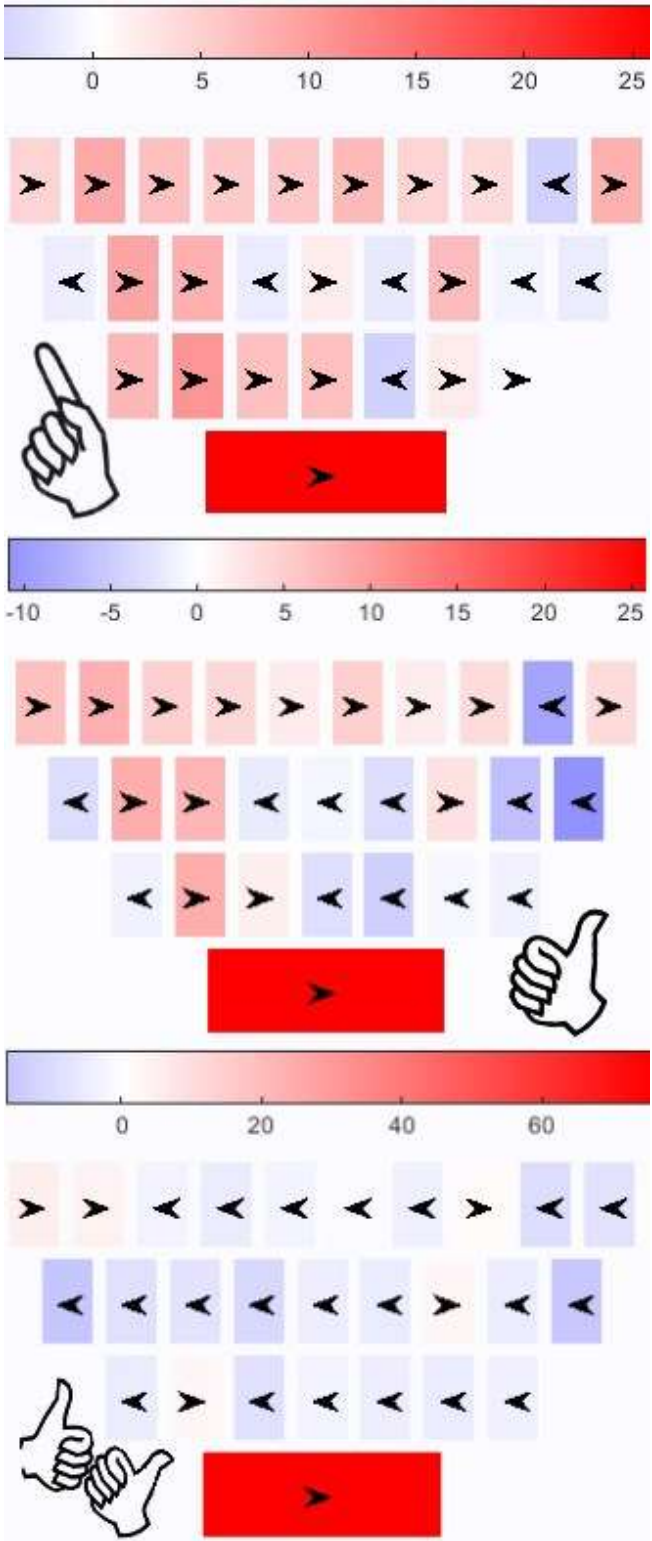

Figure 6. Horizontal Offsets Using Touch Point Method

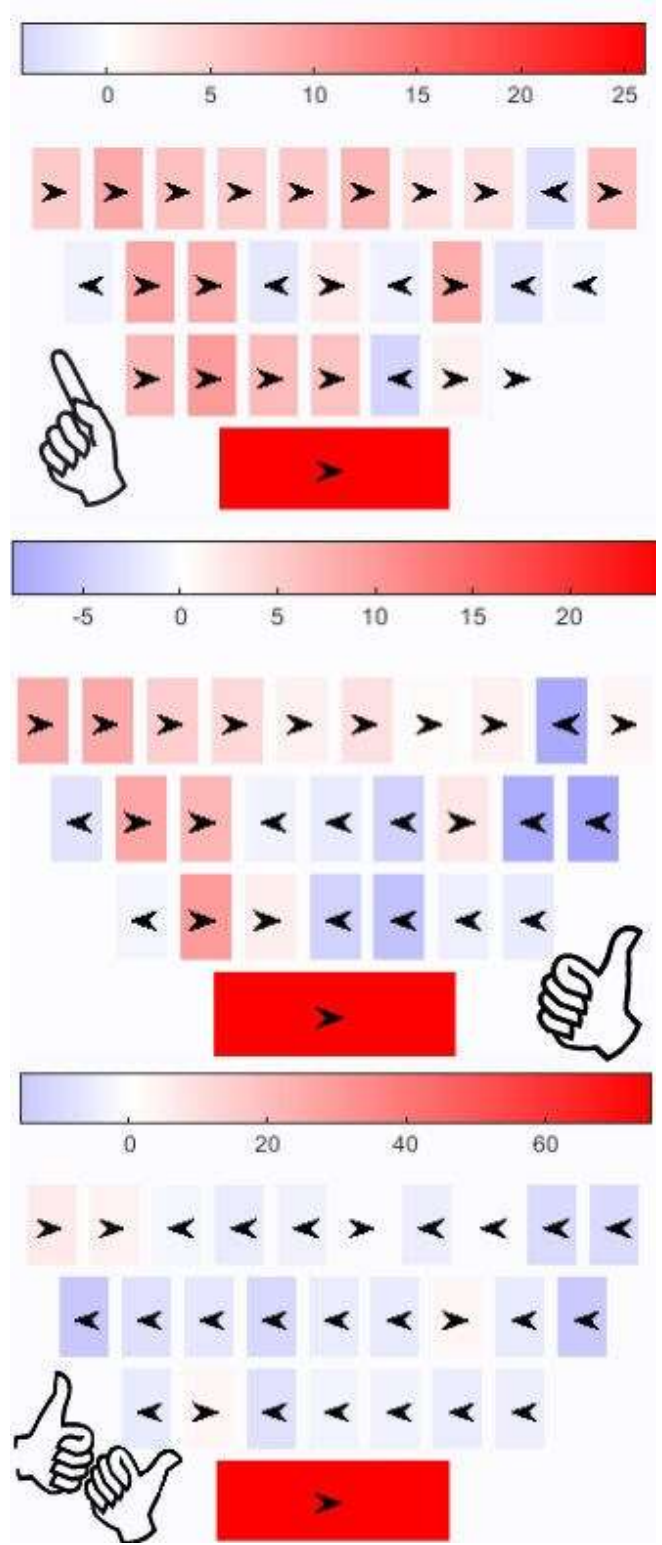

Figure 7. Horizontal Offsets Using Touch Area Method

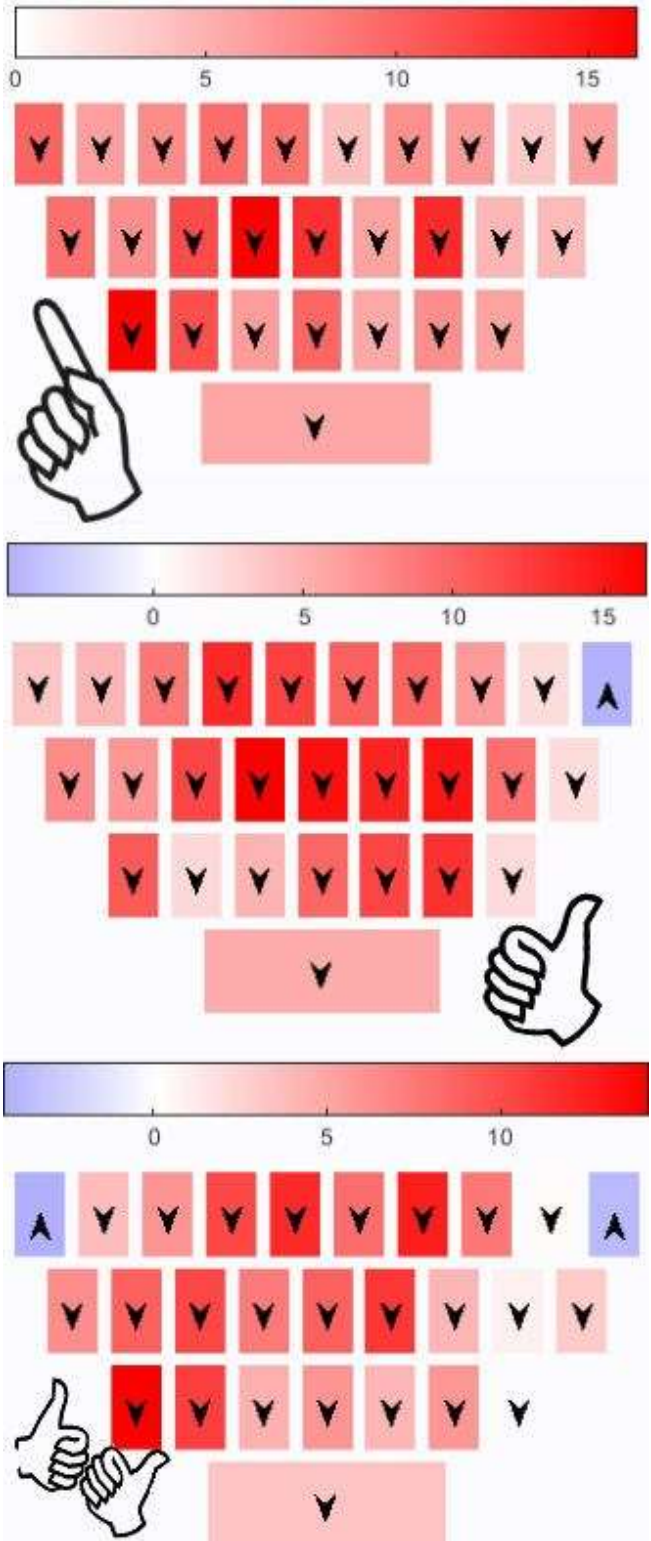

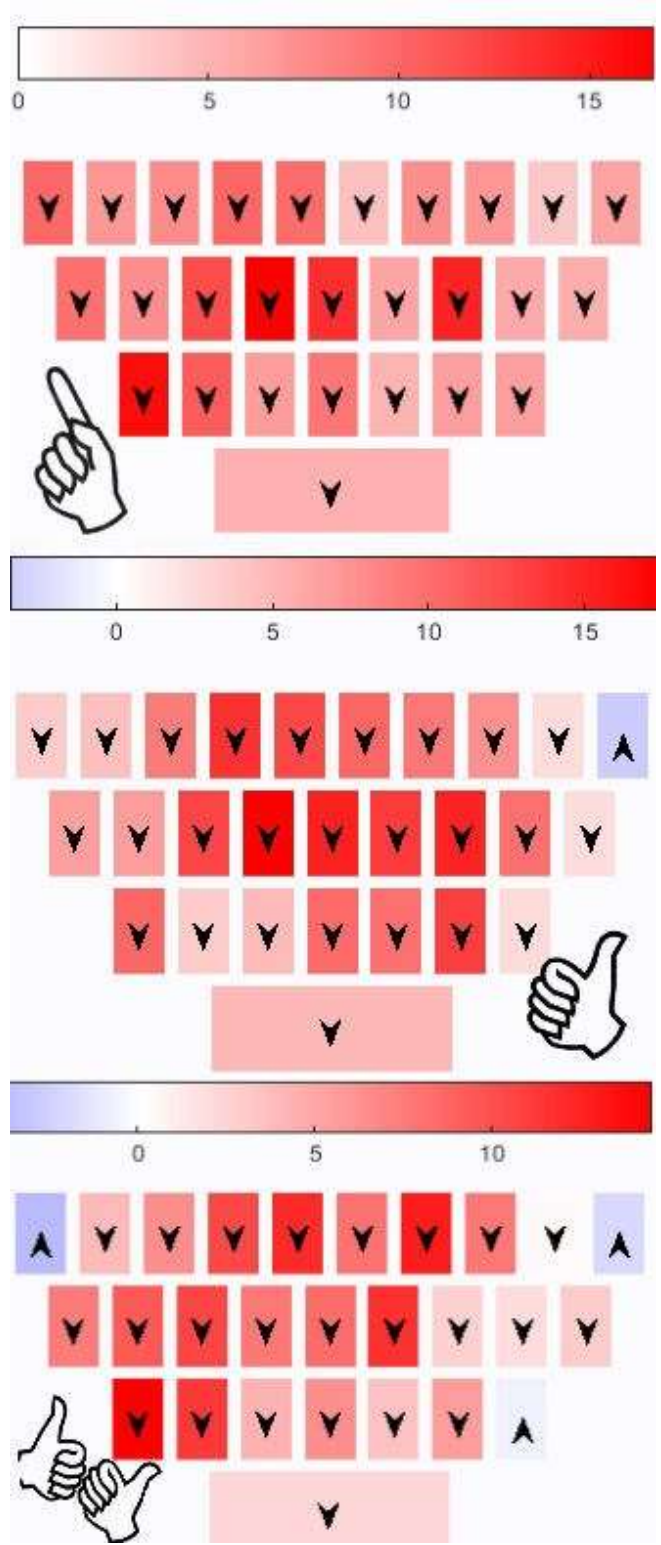

Figure 8. Vertical Offsets Using Touch Point Method

Figure 9. Vertical Offsets Using Touch Area Method

**FUTURE WORK**

During the course of our study, we identified numerous opportunities for future work. Perhaps one of the most interesting approaches (and relatively easy to implement) would be to develop a simple error correction model based on touch area and compare it to one based on touch points. To do this, one would simply learn the horizontal and vertical offsets for each key (as we have already done). Using these offsets, the overall touch point (which is what the API seems to use to determine which button was pressed) could be adjusted accordingly. These offsets may cause the touch point to shift closer to a different letter. As such, the user's input letter could be corrected to the other letter. Error rates could be computed for unadjusted versus adjusted touch points, using offset models based on touch areas and touch points. A more robust error correction model may take into account common overlaps as well, as seen in figures 4 and 5.

Another intriguing possibility would be to break the offsets up even further, based on which direction the subject's hand was coming from. For instance, if the subject just typed a "k" and then presses an "a", the subject may overshoot (or undershoot). Thus, the amount of offset and direction may also depend on where the hand is coming from. If significant improvements are found when considering touch direction, more accurate corrections may be possible.

Finally, we believe that further testing under more realistic conditions could refine touch behavior models even more. For instance, subjects driving or texting in class who are not looking at their phones continuously are more likely to make errors. Investigating how factors other than posture affect offsets could prove fruitful. Also, our study only considered portrait mode. Touch behaviors in landscape mode may prove significantly different because of differences in hand posture and thus deserves further study.

**CONCLUSION**

In this paper, we replicated and extended the work of Azenkot et al. [1] by considering the entire area touched by the subject's finger rather than just considering a touch as a single point. We showed that calculations involving touch area yield statistically significant differences from those based on touch point. For the most part, however, touch area and touch point calculations yielded similar results.

Perhaps the most insightful revelation of this work was that the touch area visualization revealed many more overlapping regions than those produced by the touch point visualization. This suggests that Azenkot et al.'s original claims about relatively low overlap are specious because of their failure to consider the entire touch area. Additionally, we showed that even though offset differences between the touch point and touch area models were fairly slight, the vertical offset differences were statistically significant for two of the postures. Though our work largely reinforced the findings of the original paper, it also suggests that testing error-correction models based on touch area may be promising, as the overlap information is more detailed. A study based on error-rate reduction using touch point and touch area error-correction models may be the only way to conclusively determine whether the observed differences between the models are meaningful.